\documentstyle[procl]{article}

\input{psfig.sty}

\def\Journal#1#2#3#4{(#1) {#2} {\bf #3}, #4}

\def\AAp{\em Astron. Astrophys.}
\def\AJ{\em Astron.~J.}
\def\ApJ{\em Astrophys.~J.}

\def\MNRAS{\em Mon. Not. R.~Astron. Soc.}
\def\Nat{\em Nature\/}

\begin{document}

\title{Gamma-rays from Collisions of Compact Objects with AGN Jets?} 

\author{W\l odek Bednarek}

\address{Department of Experimental Physics, University of \L\'od\'z,\\ 
ul. Pomorska 149/153, 90--236 {\L \'od\'z}, Poland}

\maketitle

\abstract{We consider different scenarios of collisions of compact 
objects (clouds, massive stars, supernova shock waves, or young pulsars) 
with jets in active galactic nuclei. The purpose is to find out if such
collisions can  become plausible explanations for the gamma-ray 
production in blazars. We conclude that the relativistic proton 
beam - cloud
collision scenario has problems with explanation of the
$\gamma$-ray spectrum in the observed energy range from blazars. As a 
result of collisions of massive stars, supernova shock waves, or  young
pulsars with the jet plasma, a highly oblique shocks should be formed. 
These shocks can accelerate electrons, protons to energies high
enough for production of $\gamma$-ray photons observed in blazars.
However, in order to produce observable $\gamma$-ray fluxies, the 
particles should be accelerated strongly anisotropicly, which may be the 
case of highly oblique relativistic shocks in jets of AGNs.}

\section{Introduction}

Gamma-ray production in blazars is usually interpreted in terms of a 
shock in 
the jet model. It is argued that relativistic shock moving along the jet
accelerates electrons and/or protons, creating a blob of very energetic 
particles. These particles can then interact with the soft radiation
produced in the blob itself by electrons, or with the radiation coming 
from the regions surrounding the jet (accretion disk, matter distributed 
around the jet). However some recent observations put some new light on the 
$\gamma$-ray blazars. It seems difficult to explain them in terms of this
standard  model. For example, VLBI multiple imaging of TeV BL Lac, Mrk 421,
during  1994 - 1997 (30 images) show that 3 inner components moves only
with subluminal speeds between projected linear distance 0.3 - 
3pc~\cite{pi99}. The conclusion that the blobs do not move relativisticly
at such distances is the most likely interpretation of these observations. 
Moreover, another TeV $\gamma$-ray BL Lac, Mrk 501, show the base level 
X-ray and TeV
$\gamma$-ray emission during its few months high activity period
in 1997 \cite{qu99}. Such emission is difficult to explain by a single
shock in the jet scenario, but it would rather require continuous injection 
of shocks  by the central engine into the jet.

Therefore we think that the investigation of other scenarios, in which
production of $\gamma$-rays can occur efficiently, is justified. 
In one type of these non-standard models, $\gamma$-rays are produced in 
a cascade initiated
by extremely relativistic particles accelerated close to the disk surface (see
e.g.~\cite{b97,bp99}), or by particles accelerated in almost rectilinear
reconnection regions inside the jet (e.g. \cite{bk95,bkm96,b98}).
Other models predict production of $\gamma$-rays in collisions of
hadronic beam with the clouds entering the jet\cite{bb99} or
as a result of interaction of relativistic jet with compact objects, i.e.
massive stars\cite{bp97}, supernova front waves, or very young pulsars.
In this paper we discuss these last type of models.

\section{Collision of a cloud with the hadronic beam}

It is possible that AGN jets contain relativistic hadronic beams
propagating along the jet axis. Such
beams of particles may be accelerated by the large scale electric fields
generated by  rotating accretion disks or  black holes in a perpendicular
magnetic field\cite{Lov76,bz77}, by a magnetic reconnection occurring on or
close to the disk surface~\cite{hts92}, by magnetic reconnection occuring 
inside the jet~\cite{lp92}, or by highly oblique  shocks present in the 
jet. It seems likely that the Broad Line Region clouds  (BLR) enter
frequently  into the region of the jet. Such clouds, they are probably
extended atmospheres of massive stars, have typical densities 
$\sim 10^{10} - 10^{12}$ cm$^{-2}$ and dimensions $\sim 10^{12} - 
10^{13}$ cm. They can create
significant target for  relativistic hadronic beam. As a result, the
distant observer located at the direction of the beam should detect 
variable $\gamma$-rays.   
The hadronic beam BLR cloud interaction model has
been investigated even before the discovery of $\gamma$-ray emission from
blazars (see e.g.  Rose et al.~\cite{retal84}). It has been recently 
explored as a possible explanation of the blazar phenomenon by 
Dar \& Laor~\cite{dl97} and Beall \& Bednarek~\cite{bb99}. 

In our paper\cite{bb99}, we show that the energy losses of a hadronic 
beam on the excitation of plasma waves in the cloud dominate over pion
production losses if hadrons have Lorentz factors below a few  hundred. 
Therefore, the $\gamma$-ray spectra below $\sim
30$ GeV should be suppressed. This feature is in fact required by  the
observations of blazars which usually show lower fluxies at lower energy
$\gamma$-rays and by the lack of strong variability at these lower 
energies (see e.g. the spectra of Mrk 421 and Mrk 501). However, the
$\gamma$-ray fluxies expected in such a scenario below $\sim 300$ MeV 
are too
low. It is difficult to find  another mechanism which contributes
significantly to this energy range. Thus, we conclude that hadronic beam -
cloud interaction model has problems with explanation of the $\gamma$-ray
spectrum observed from blazars at energies below $\sim 300$ MeV.

\section{Collision of a massive star with the jet}

It is expected that active galactic nuclei are surrounded by huge stellar
clusters containing $10^6 - 10^9$ stars.  For example, observations of 
nearby relatively small galaxy, M32, show central stellar density of 
$\sim 2\times 10^5$ M$_{\odot}$ pc$^{-3}$ within $\sim 1$ pc~\cite{la92}.
Probability of stellar destruction as a result of star collisions depends on
the escape velocity from the region containing the main mass of the star, on
the  radius of the star and on stellar density. The massive stars, with
higher escape velocity, collide less frequently than e.g. Solar type 
stars. Simulations of the evolution of very massive stellar clusters with 
the black hole inside, show that significant amount of massive stars,
instead of colliding, finishes  their life as a supernovae~\cite{ds83}.     

The massive stars, of the Wolf-Rayet (WR) and OB type, are characterized 
by very strong  stellar winds. The high energy processes during 
interaction of
such winds with the relativistic plasma of the jet has been recently
investigated by Bednarek \& Protheroe~\cite{bp97}. The pressure of the stellar
wind can be estimated from
\begin{equation}
P_{\rm w} = Mv_\infty/4\pi r^2\approx 1.6\times 10^4M_{-5}v_3/R_{12}^{2}
(r/R)^{2} {\rm ~erg~cm}^{-3},
\label{eq:swind}
\end{equation}
where the mass-loss rate of the star is $M = 10^{-5}M_{-5}$ M$_{\odot}$
yr$^{-1}$, the wind velocity is $v_{\infty} = 3\times 10^8v_3$ cm 
s$^{-1}$, $R = 10^{12}R_{12}$ cm is the radius of the star, and $r$ is the
distance from the star surface.
The wind pressure is balanced by the ram pressure of the jet plasma
\begin{equation}
P_{\rm j} = L_j/\pi c\theta^2l^2\approx 15L_{45}/\theta_5^{2}l_{0.1}^{2}
{\rm ~erg~cm}^{-3}, 
\label{eq:starwind}
\end{equation}
where $L_{45}$ is the jet power in units of $10^{45}$ erg s$^{-1}$,
$\theta_5$ is its opening angle in units of $5^{\rm o}$, and $l_{0.1}$ 
is the distance in 0.1 parsec from the base of the jet. As a result of 
this interaction  a double shock structure is formed at the distance 
from the star
\begin{equation}
r/R\approx 33M_{-5}^{1/2}v_3^{1/2}\theta_5l_{0.1}/R_{12}L_{45}^{1/2}.
\label{eq:locshock}
\end{equation}
It is shown\cite{bp97}, that such standing shock can accelerate 
electrons to high energies. These electrons can produce $\gamma$-ray 
photons by scattering  thermal radiation coming from a massive star, and
synchrotron X-ray  photons in the magnetic field supplied by the star.
However, observable $\gamma$-ray fluxies can be produced in such a model if
the acceleration of electrons is highly anisotropic. Such situation might
happen  if electrons are drifting along the surface of the oblique shock (so
called shock drift acceleration on  the superluminal shocks). We estimated
that many stars may be found inside the jet at the same moment which
significantly increases the probability of detection of the $\gamma$-rays. The
movement of the star through the jet and disturbances present in the jet and
the stellar wind, which create instability in the location of the
shock, can be responsible of strong variability of $\gamma$-ray emission from
blazars.

\section{Collision of a supernova shock with the jet}

As we noted above,  big number of massive stars in galactic nuclei should
finish their life as a supernovae. The expending supernova shell can
significantly  perturb the plasma flow in the jet if the pressure of 
material in the supernova front wave is comparable to the pressure of the
plasma in the jet. The pressure of the supernova front wave can be 
estimated from 
\begin{equation}
P_{\rm SN} = L_{\rm SN}/V_{\rm SN}\approx 9\times 10^{-3} L_{51}/
r_{0.1}^{3} {\rm ~erg ~cm}^{-3},
\label{eq:pressn}
\end{equation}
where $L_{51}$ is the supernova kinetic power, $L_{\rm SN}$, in units of
$10^{51}$ erg, and $r_{0.1}$ is the radius of the volume, $V_{\rm SN}$, 
occupied by the expending supernova in units of 0.1 pc. By comparing 
supernova shell pressure with the jet plasma pressure 
(Eq.~\ref{eq:starwind}), we can
estimate  the distance from the jet at which the expending shell can 
perturb significantly the jet
\begin{equation}
r_{0.1}\approx 3.5\times 10^{-2}(L_{51}\theta_5^2\l_{0.1}^2/L_{45})^{1/3}.
\label{eq:distsnjet}
\end{equation}
It is clear that supernovae have to explode relatively close to the jet 
or have to find itself close to  the jet at a later time after explosion 
as a result of  initial fast motion of the presupernova star around the
galactic nuclei. If the supernova is energetic enough, it may even 
completely obstruct the jet plasma flow for some time. It happens when 
the radius of the expending supernova becomes comparable to the 
perpendicular extend of the jet defined by the jet opening angle and 
the distance from the base of the jet, i.e. $r_{0.1} = \theta_5l_{0.1}$. 
This condition is fulfilled if 
\begin{equation}
L_{51} \ge 1.5\times 10^3L_{45}\theta_5l_{0.1}.
\label{eq:blok}
\end{equation}
Therefore it may happen only for less powerful jets and at small
distances from the base of the jet.

We suppose that the large scale shocks created by the supernova shock 
waves in the jet, may be much efficient accelerators of particles to 
very high energies than the classical supernova shock waves because of 
much stronger magnetic field strength in the shock region. The magnetic 
field strength in the jet, can be estimated if we assume that the magnetic
energy density is in equipartition with the radiation energy density 
close to the base of the jet. From the observed UV power in the case of 
$\gamma$-ray blazar 3C 273 ($\sim 3\times 10^{46}$ erg 
s$^{-1}$~\cite{li95}), we can determine the radiation energy density and 
so the magnetic field strength at the base of the jet. Assuming that the
magnetic field in the jet drops inversely proportionally to the distance 
from the base of the jet $l$ and that the shock created in the jet at the
distance $l$ 
extends far away along the jet, we estimate the maximum possible 
energies reached by  particles with the charge $Z$ on~\cite{bed98} 
\begin{equation}
E_{\rm Z, max}\approx 10^{13} \chi Z/(1 + l/r_{\rm in}) {\rm ~GeV},
\label{eq:maksen}
\end{equation}
where $\chi$ is the acceleration coefficient (see next section for the
expected value of $\chi$), and $r_{\rm in}\approx 9\times 10^{14}$ cm is 
the inner radius of 
the disk in 3C 273 ~\cite{li95}. These energies can be comparable to
the highest energies observed  in the cosmic rays.

\section{Collision of a very young pulsar with the jet}

It is expected that in some explosions of massive stars also very young
pulsars are formatted. The relativistic wind, produced by such very young 
pulsar, exerts a pressure 
\begin{equation}
P_{\rm P}= L_{\rm P}/4\pi r^2c \approx  8.8\times 10^{-4}
B_{12}^2/P_{\rm ms}^4r_{0.1}^2 {\rm ~erg ~cm}^{-3},
\label{eq:presps}
\end{equation}
where the rotational energy loss rate by the pulsar in the form of 
the wind is $L_{\rm P}\cong 3\times 10^{43} B_{12}^2P_{\rm ms}^{-4}$
erg s$^{-1}$, and $B_{12}$ and $P_{\rm ms}$ are the surface magnetic 
field and the period of the neutron  star in units of $10^{12}$ G and
miliseconds, respectively. $r_{0.1}$ is the radius of the volume in which
relativistic wind is confined in units of 0.1pc, and $c$ is the velocity
of light. If such a pulsar find  itself inside the jet then as a result 
of the pulsar wind - jet plasma interaction a shock structure forms with 
the characteristic radius which can be estimated by comparison of
Eqs.~(\ref{eq:starwind}), and~(\ref{eq:presps})  
\begin{equation}
r_{0.1}\approx 7.7\times 10^{-3} B_{12}\theta_5l_{0.1}/P_{\rm ms}^2
L_{\rm 45}^{1/2}. 
\label{eq:radpsjet}
\end{equation}
We can also estimate the strength of the magnetic field in the shock 
region assuming that it is of a dipole type inside the pulsar 
magnetosphere
and drops as $r^{-1}$ with the distance $r$ in the pulsar wind zone.
It is equal to
\begin{equation}
B_{\rm sh}\approx 0.14B_{12}/P_{\rm ms}^2r_{0.1} {\rm ~G}.
\label{eq:magshock}
\end{equation}
Using Eq.~(\ref{eq:radpsjet}) for the shock dimension, we obtain
\begin{equation}
B_{\rm sh}\approx 18L_{45}^{1/2}/\theta_5l_{0.1} {\rm ~G}.
\label{eq:magshock2}
\end{equation}
It is interesting that the value of the magnetic field in the shock 
region do not depend on the parameters of the pulsar but only on the 
parameters of the jet! The only condition which has to be fulfilled is 
that the jet pressure has to be balanced by the pulsar pressure in the 
pulsar wind zone. This happens for the condition 
\begin{equation}
B_{12}/P_{\rm ms}^3 > 2\times 10^{-9}L_{45}^{1/2}/\theta_5l_{0.1}.
\label{eq:dippsjet}
\end{equation}
The particles accelerated in the shock region, where the magnetic field 
strength is given by Eq.~(\ref{eq:magshock2}), gains energy at a rate 
\begin{equation}
E = \chi ZecB_{\rm sh} {\rm ~erg ~s}^{-1}.
\label{eq:engain}
\end{equation}
The maximum energies of electrons are limited by the synchrotron losses. 
In the absence of other losses, electrons can reach energies
\begin{equation}
E_{\rm max} = 6\times 10^4(\chi /B_{\rm sh})^{1/2} {\rm ~GeV}.
\label{eq:emax}
\end{equation}
The maximum energies of synchrotron photons, produced by these electrons,
are
\begin{equation}
\varepsilon_{\rm x}\approx (B_{\rm sh}/B_{\rm cr})(E_{\rm max}^2
/m_{\rm e}),
\label{eq:eps}
\end{equation}
where $B_{\rm cr} = 4.4\times 10^{14}$ G. If the synchrotron
photons with energies observed from the jet of Mrk 501, 
$\varepsilon_{\rm x} = 2\times 10^{-4}$ GeV,
are produced by such electrons, then we can estimate the acceleration
efficiency  $\chi$ using Eqs.~(\ref{eq:emax}), and~(\ref{eq:eps}),
\begin{equation}
\chi\approx 2.8\times 10^{-10}m_{\rm e} B_{\rm cr}\varepsilon_{\rm x},
\label{eq:accef}
\end{equation}
which is $\chi\approx 10^{-3}$ for the above value of 
$\varepsilon_{\rm x}$.

If the maximum energies of accelerated protons are limited only by
their synchrotron losses, then protons can reach energies as high as 
\begin{equation}
E_{\rm p,max} = 2\times 10^{11}(\chi /B_{\rm sh})^{1/2} {\rm ~GeV}.
\label{eq:pmax}
\end{equation}
For the value of $\chi$, estimated above, and the value of the magnetic
field at the shock $\sim 10$ G, obtained for typical parameters of the
jet (Eq.~\ref{eq:magshock2}), protons can reach energies  as high as 
$\sim 2\times 10^9$ GeV. These protons still fulfill the condition that 
the dimension of the shock structure should be smaller  than the Larmor 
radius of protons. However the maximum energies of protons can be
limited by their energy losses in collisions with synchrotron photons 
which are produced by accelerated electrons. The decay products of 
pions, created in proton-photon collisions (electrons, positrons and 
very high energy $\gamma$-rays), initiate cascade in the magnetic field. 
Such mechanism has been proposed  as a possible explanation of blazar
phenomenon (so called  synchrotron-proton
blazar model~\cite{mb92,mann93,mu99}). However the original version of 
this 
model requires magnetic field of the order of a few tens of Gauss in the  blob
moving relativisticly along the jet in order to accelerate protons on
sufficiently short time scale. Such magnetic fields seem to be unacceptable
at larger distances from the accretion disk. However, as we  show here, the
magnetic fields of the required order should be present  in the shock region
if young pulsar collides with the relativistic jet. Although the shock
considered by us is stationary in the jet frame, the relativistic flow of
plasma through such highly oblique shock should result in strong collimation
of accelerated electrons and protons, which is in  fact equivalent to the
situation that the shock moves relativisticly  through the jet. The radiation
produced by electrons and protons will be strongly collimated along the shock
surface, i.e. also along the jet axis.

\section*{Acknowledgments}

The present work was supported by the {\em Komitet Bada\'n Naukowych}   
through the grant 2P03D 001 14.

\end{document}